\documentclass[adraft]{eptcs}

\title{A better model of computation for digital physics?}
\author{Anton Salikhmetov
\email{anton.salikhmetov@gmail.com}}
\def\titlerunning{A better model of computation for digital physics?}
\def\authorrunning{A. Salikhmetov}

\usepackage[utf8]{inputenc}
\usepackage[english]{babel}
\usepackage[title]{appendix}
\usepackage{amsmath}
\usepackage{amssymb}
\usepackage{amsthm}
\usepackage{fixmath}
\usepackage{wrapfig}
\usepackage{caption}
\usepackage{subcaption}
\usepackage{tikz}
\usepackage{tikz-inet}

\DeclareFontFamily{U}{mathb}{\hyphenchar\font45}
\DeclareFontShape{U}{mathb}{m}{n}{
<-6> mathb5
<6-7> mathb6
<7-8> mathb7
<8-9> mathb8
<9-10> mathb9
<10-12> mathb10
<12-> mathb12
}{}
\DeclareSymbolFont{mathb}{U}{mathb}{m}{n}
\DeclareMathSymbol{\righttoleftarrow}{\mathrel}{mathb}{"FD}

\tikzset{every node/.style = {node distance=0em, scale=0.8}}

\newcommand{\ar}{\text{\textsf{ar}}}

\begin{document}
\maketitle

\begin{abstract}
This note is meant to invite the reader to consider interaction nets, a relatively recently discovered model of computation, as a possible alternative for cellular automata which are often employed as the basis for digital physics.
Defined as graph-like structures (in contrast to the grids for cellular automata), interaction nets possess a set of interesting properties, such as locality, linearity, and strong confluence, which together result in so-called clockless computation in the sense that they do not require any global clock in order to operate.
We believe that an attempt of using interaction nets as a replacement for cellular automata may lead to a new view in digital physics.
\end{abstract}

\section{Introduction}

The idea of representing the world as a process of executing a digital computer program seems to have been circulating since about 1960s.
In recent works~\cite{hooft} cellular automata appear to have been the most widely used (if not the only one) model of computation put forward as the basis for modeling such a digital computer.
Traditionally, the grid of a cellular automaton is seen as representation of space, and the sequence of its states as representation of time.

In computer science, cellular automata is one member in quite a big array of models of computation.
Models of computation are equivalent in the sense that any solvable problem can be solved within those models.
However, properties which models possess can be very different.
In this note we would like to advertise \textit{interaction nets} which we believe might possibly lead to a new view in digital physics.

Interaction nets are one of graphical models of computation based on the notion of ``computation as interaction''.
They were devised by Yves Lafont~\cite{lafont} as a generalisation of the proof structures of linear logic.
This model of computation benefits from the following properties:
\begin{itemize}
\item \textbf{locality} in the sense that only two adjacent nodes can interact at a time, and each such step of computation is completely independent from the rest of a net;
\item \textbf{linearity} in the sense that each step of computation can be done in constant time, therefore total computation time is linear on the number of steps;
\item \textbf{strong confluence} in the sense that the order in which the steps of computation are performed does not influence the process of computation.
\end{itemize}

Let us make some remarks regarding representation of space and time with interaction nets.
First, as arbitrary graph-like structures, interaction nets are inherently three-dimensional, unlike cellular automata where one has to assign dimensionality to the system, perhaps artificially.
Second, while causality does take place in interaction nets just like for cellular automata, the notion of a global state is rather irrelevant due to strong confluence, so computing with interaction nets is essentially clockless.

The next section first gives a short overview of interaction nets and then discusses their properties more formally among some other aspects.
For a more thorough introduction as well as examples of how exactly interaction nets can perform actual computations, we urge the reader to follow~\cite{models}.

\clearpage
\section{Interaction Nets}

This section gives a rather informal brief introduction to interaction nets and their textual representation called the interaction calculus~\cite{calculus}.
Here, we intentionally omit the notion of \textit{interface} which is specific to some applications but is not crucial for this model of computation.

\begin{figure}
\centering
\caption{Primitives}
\label{primfig}
\begin{subfigure}[t]{0.3\textwidth}
\centering
\caption{Agent}
\label{primagentfig}
$$
\begin{tikzpicture}[baseline=(i.base)]
\inetcell(a){$\phantom X\alpha\phantom X$}[R]
\inetwirefree(a.pal)
\inetwirefree(a.left pax)
\inetwirefree(a.right pax)
\node (0) [right=of a.above pal] {$x_0$};
\node (1) [left=of a.above left pax] {$x_1$};
\node (i) [left=of a.above middle pax] {$\vdots$};
\node (n) [left=of a.above right pax] {$x_n$};
\end{tikzpicture}
$$
\end{subfigure}
\hfill
\begin{subfigure}[t]{0.2\textwidth}
\centering
\caption{Wiring}
\label{primwirefig}
$$
\begin{tikzpicture}[baseline=(w)]
\matrix[row sep=1.5em]{
\node (l) {$x_1$}; &
\node (i) {$\cdots$}; &
\node (r) {$x_{2k}$}; \\
\node (l') {$\phantom{x_1}$}; &
\node (w) {$\omega$}; &
\node (r') {$\phantom{x_{2k}}$}; \\
};
\inetbox{(l') (w) (r')}(b)
\inetwirecoords(l)(intersection cs:
first line={(b.north west)--(b.north east)},
second line={(l)--(l')})
\inetwirecoords(r)(intersection cs:
first line={(b.north west)--(b.north east)},
second line={(r)--(r')})
\end{tikzpicture}
$$
\end{subfigure}
\hfill
\begin{subfigure}[t]{0.4\textwidth}
\centering
\caption{Tree}
\label{primtreefig}
$$
\begin{tikzpicture}[baseline=(a)]
\inetcell[double](a){$\phantom x t\phantom x$}[U]
\inetwirefree(a.pal)
\inetwirefree(a.left pax)
\inetwirefree(a.right pax)
\node [below=of a.middle pax] {$\cdots$};
\end{tikzpicture}
\equiv
\begin{tikzpicture}[baseline=(a)]
\matrix[row sep=2.5em]{
\inetcell(a){$\phantom x \alpha \phantom x$}[U] \\
\node (i) {$\cdots$} ; \\ };
\inetcell[double, left=of i](t1){$\ t_1\ $}[U]
\inetcell[double, right=of i](tn){$\ t_n\ $}[U]
\inetwirefree(a.pal)
\inetwire(a.left pax)(t1.pal)
\inetwire(a.right pax)(tn.pal)
\inetwirefree(t1.left pax)
\inetwirefree(t1.right pax)
\node [below=of t1.middle pax] {$\cdots$};
\node (it) [below=of a.middle pax] {$\cdots$};
\inetwirefree(tn.left pax)
\inetwirefree(tn.right pax)
\node [below=of tn.middle pax] {$\cdots$};
\end{tikzpicture}
\text{ or }
\begin{tikzpicture}[baseline=(a)]
\inetcell[double,opacity=0](a){$t$}[U]
\inetwirecoords(a.above pal)(a.middle pax)
\node (i) [below=of a.middle pax] {$x$};
\end{tikzpicture}
$$
\end{subfigure}
\end{figure}
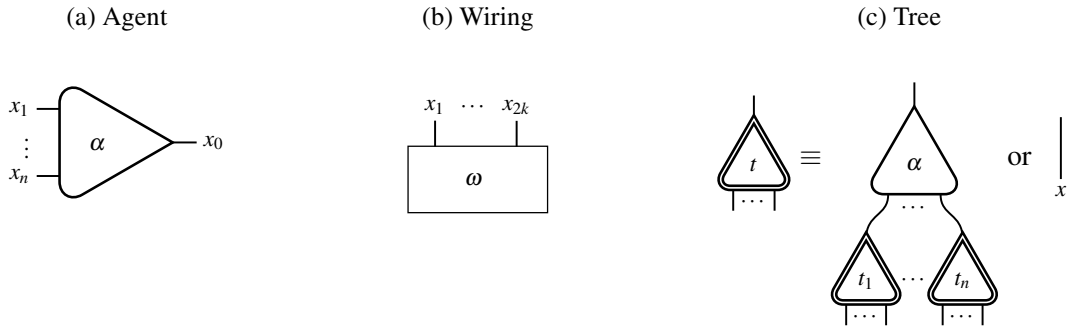

Interaction nets are graph-like structures consisting of primitives shown in Figure~\ref{primfig}.
\textit{Agents} of type $\alpha$ can be graphically represented as shown in Figure~\ref{primagentfig}.
Agents have \textit{arity} $\ar(\alpha) \ge 0$.
If $\ar(\alpha) = n$, the agent $\alpha$ has $n$ \textit{auxiliary ports} $x_1,\dots, x_n$ in addition to its \textit{principal port} $x_0$.
All agent types belong to a set $\Sigma$ called \textit{signature}.
Any port must be connected to exactly one edge.
\textit{Wiring} $\omega$ on Figure~\ref{primwirefig} consists solely of edges.
Inductively defined \textit{trees} on  Figure~\ref{primtreefig} correspond to \textit{terms} $t ::= \alpha(t_1,\dots, t_n)\ |\ x$ in the interaction calculus, where $x$ is called a \textit{name}.

Any net $N$ can be redrawn using the previously defined wiring and tree primitives as follows:
$$
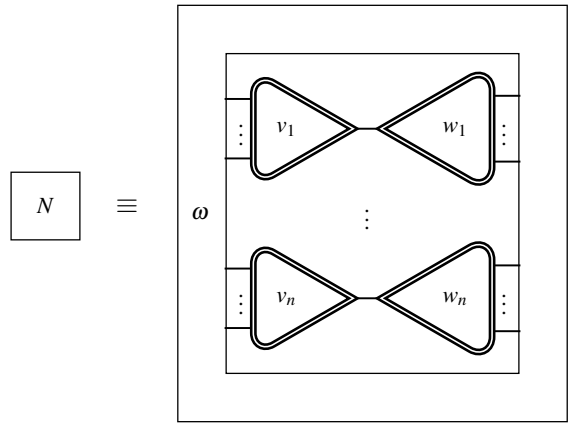
\begin{tikzpicture}[baseline=(n.base)]
\node (n) {$N$};
\inetbox{(n)}(b)
\end{tikzpicture}
\quad \equiv \quad
\begin{tikzpicture}[baseline=(wl)]
\matrix{
\node (wnwo) {}; & & & & & & & \\
\node {$\phantom\omega$}; &
\node (wnw) {$\phantom\omega$}; & & & & \\
& & &
\inetcell[double](v1){$\ \ v_1\ \ $}[R] & &
\inetcell[double](w1){$\ \ w_1\ \ $}[L] & & \\
\node (wl) {$\omega$}; & & & &
\node (ei) {$\vdots$}; & & &
\node (wr) {$\phantom\omega$}; \\
& & &
\inetcell[double](vn){$\ \ v_n\ \ $}[R] & &
\inetcell[double](wn){$\ \ w_n\ \ $}[L] & \\
& & & & & & \node (wse) {$\phantom\omega$}; & \\
& & & & & & & \node (wseo) {$\phantom\omega$}; \\
};
\inetbox[inner sep=3em]{(wnw) (wse)}(bo)
\inetbox{(wnw) (wse)}(bi)
\inetwire(v1.pal)(w1.pal)
\inetwire(vn.pal)(wn.pal)
\node [left=of v1.middle pax] {$\vdots$};
\node [left=of vn.middle pax] {$\vdots$};
\node [right=of w1.middle pax] {$\vdots$};
\node [right=of wn.middle pax] {$\vdots$};
\inetwirecoords(w1.right pax)(intersection cs:
first line={(bi.north east)--(bi.south east)},
second line={(w1.right pax)--(w1.above right pax)})
\inetwirecoords(w1.left pax)(intersection cs:
first line={(bi.north east)--(bi.south east)},
second line={(w1.left pax)--(w1.above left pax)})
\inetwirecoords(wn.right pax)(intersection cs:
first line={(bi.north east)--(bi.south east)},
second line={(wn.right pax)--(wn.above right pax)})
\inetwirecoords(wn.left pax)(intersection cs:
first line={(bi.north east)--(bi.south east)},
second line={(wn.left pax)--(wn.above left pax)})
\inetwirecoords(v1.right pax)(intersection cs:
first line={(bi.north west)--(bi.south west)},
second line={(v1.right pax)--(v1.above right pax)})
\inetwirecoords(v1.left pax)(intersection cs:
first line={(bi.north west)--(bi.south west)},
second line={(v1.left pax)--(v1.above left pax)})
\inetwirecoords(vn.right pax)(intersection cs:
first line={(bi.north west)--(bi.south west)},
second line={(vn.right pax)--(vn.above right pax)})
\inetwirecoords(vn.left pax)(intersection cs:
first line={(bi.north west)--(bi.south west)},
second line={(vn.left pax)--(vn.above left pax)})
\end{tikzpicture}
$$
which in the interaction calculus corresponds to a \textit{configuration} $\langle v_1 = w_1, \dots,\ v_n = w_n \rangle$ which is an unordered multiset of \textit{equations} $v_i = w_i$, where $v_i$ and $w_i$ are arbitrary terms.
The wiring $\omega$ translates to names, and each name has to occur exactly twice in a configuration.

For configurations, so-called \textit{$\alpha$-conversion} is defined as follows: both occurrences of any name can be replaced with any new name if the latter does not occur in a given configuration.
Configurations are considered equal up to $\alpha$-conversion.

\clearpage
When two agents are connected to each other with their principal ports, they form an \textit{active pair}.
For active pairs one can introduce \textit{interaction rules} which describe how the active pair rewrites to another net.
Graphically, any interaction rule can be represented as follows:
$$
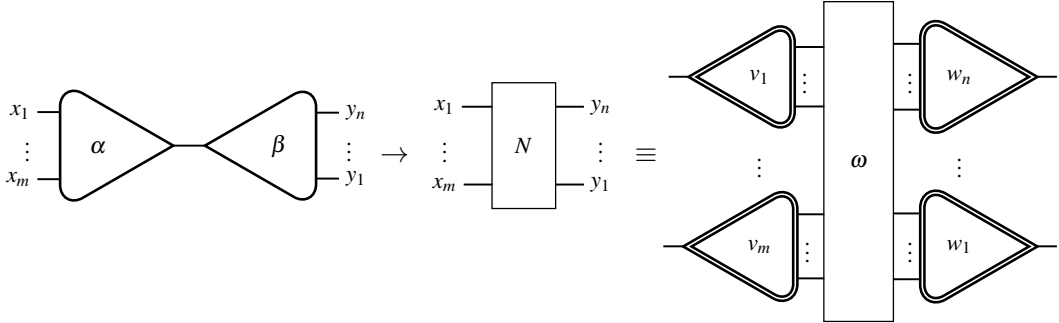
\begin{tikzpicture}[baseline=(yi.base)]
\matrix[column sep=1em]{
\inetcell(a){$\phantom X\alpha\phantom X$}[R] &
\inetcell(b){$\phantom X\beta\phantom X$}[L] \\ };
\inetwirefree(a.left pax)
\inetwirefree(a.right pax)
\inetwire(a.pal)(b.pal)
\inetwirefree(b.left pax)
\inetwirefree(b.right pax)
\node (x1) [left=of a.above left pax] {$x_1$};
\node (xi) [left=of a.above middle pax] {$\vdots$};
\node (xn) [left=of a.above right pax] {$x_m$};
\node (y1) [right=of b.above left pax] {$y_1$};
\node (yi) [right=of b.above middle pax] {$\vdots$};
\node (yn) [right=of b.above right pax] {$y_n$};
\end{tikzpicture}
\rightarrow
\begin{tikzpicture}[baseline=(xi.base)]
\matrix[column sep=1.5em]{
\node (x1) {$x_1$}; &
\node (t) {$\phantom x$}; &
\node (yn) {$y_n$}; \\
\node (xi) {$\vdots$}; &
\node (n) {$N$}; &
\node (yi) {$\vdots$}; \\
\node (xn) {$x_m$}; &
\node (b) {$\phantom x$}; &
\node (y1) {$y_1$}; \\ };
\inetbox{(b) (t)}(box)
\inetwirecoords(x1)(intersection cs:
first line={(box.north west)--(box.south west)},
second line={(x1)--(yn)})
\inetwirecoords(xn)(intersection cs:
first line={(box.north west)--(box.south west)},
second line={(xn)--(y1)})
\inetwirecoords(yn)(intersection cs:
first line={(box.north east)--(box.south east)},
second line={(x1)--(yn)})
\inetwirecoords(y1)(intersection cs:
first line={(box.north east)--(box.south east)},
second line={(xn)--(y1)})
\end{tikzpicture}
\equiv
\begin{tikzpicture}[baseline=(wl)]
\matrix[column sep=1.5em]{
& \node (wt) {$\phantom\omega$}; & \\
\inetcell[double](t1){$\ \ v_1\ \ $}[L] & &
\inetcell[double](v1){$\ \ w_n\ \ $}[R] \\
\node (ti) {$\vdots$}; &
\node (wl) {$\omega$}; &
\node (ei) {$\vdots$}; \\
\inetcell[double](tm){$\ \ v_m\ \ $}[L] & &
\inetcell[double](vn){$\ \ w_1\ \ $}[R] \\
& \node (wb) {$\phantom\omega$}; & \\
};
\inetbox{(wt) (wb)}(b)
\inetwirefree(t1.pal)
\inetwirefree(tm.pal)
\inetwirefree(v1.pal)
\inetwirefree(vn.pal)
\node [right=of t1.middle pax] {$\vdots$};
\node [right=of tm.middle pax] {$\vdots$};
\node [left=of v1.middle pax] {$\vdots$};
\node [left=of vn.middle pax] {$\vdots$};
\inetwirecoords(t1.right pax)(intersection cs:
first line={(b.north west)--(b.south west)},
second line={(t1.right pax)--(t1.above right pax)})
\inetwirecoords(t1.left pax)(intersection cs:
first line={(b.north west)--(b.south west)},
second line={(t1.left pax)--(t1.above left pax)})
\inetwirecoords(tm.right pax)(intersection cs:
first line={(b.north west)--(b.south west)},
second line={(tm.right pax)--(tm.above right pax)})
\inetwirecoords(tm.left pax)(intersection cs:
first line={(b.north west)--(b.south west)},
second line={(tm.left pax)--(tm.above left pax)})
\inetwirecoords(v1.right pax)(intersection cs:
first line={(b.north east)--(b.south east)},
second line={(v1.right pax)--(v1.above right pax)})
\inetwirecoords(v1.left pax)(intersection cs:
first line={(b.north east)--(b.south east)},
second line={(v1.left pax)--(v1.above left pax)})
\inetwirecoords(vn.right pax)(intersection cs:
first line={(b.north east)--(b.south east)},
second line={(vn.right pax)--(vn.above right pax)})
\inetwirecoords(vn.left pax)(intersection cs:
first line={(b.north east)--(b.south east)},
second line={(vn.left pax)--(vn.above left pax)})
\end{tikzpicture}
$$
where $\alpha, \beta \in \Sigma$, and the net $N$ is redrawn using primitives of wirings and trees in order to translate the rule into the interaction calculus as $\alpha[v_1, \dots, v_m] \bowtie \beta[w_1, \dots, w_n]$ using Lafont's notation.
A net with no active pairs is said to be in \textit{normal form}.
A signature $\Sigma$ (with mapping $\ar$ defined on it) along with a set of interaction rules defined for agents $\alpha \in \Sigma$ together constitute an \textit{interaction system}.

Now, let us consider an example for the notions introduced above in this section.
Figure~\ref{egfig} shows two interaction rules for commonly used agents $\epsilon$ and $\delta$ and a simple interaction net to which these interaction rules are applied.
Using Lafont's notation, the erasing rule from Figure~\ref{egerasefig} is written as $\epsilon \bowtie \alpha[\epsilon,\dots, \epsilon]$, while the duplication rule given in Figure~\ref{egdupfig} can be represented as follows:
$$
\delta[\alpha(x_1,\dots, x_n), \alpha(y_1,\dots, y_n)] \bowtie \alpha[\delta(x_1, y_1),\dots, \delta(x_n, y_n)].
$$
Figure~\ref{egloopfig} provides an example of a non-terminating net which reduces to itself.
In terms of the interaction calculus, one can write this net as a configuration $\langle\delta(\epsilon, x) = \gamma(x, \epsilon)\rangle$.

\begin{figure}[b]
\centering
\caption{Example}
\label{egfig}
\begin{subfigure}[t]{0.3\textwidth}
\centering
\caption{Erasing}
\label{egerasefig}
$$
\begin{tikzpicture}[baseline=(a)]
\matrix[row sep=1em]{
\inetcell(e){$\epsilon$}[D] \\
\inetcell(a){$\phantom x\alpha\phantom x$}[U] \\ };
\node (i) [below=of a.middle pax] {$\dots$};
\inetwirefree(a.left pax)
\inetwirefree(a.right pax)
\inetwire(e.pal)(a.pal)
\end{tikzpicture}
\rightarrow
\begin{tikzpicture}[baseline=(i.base)]
\matrix{
\inetcell(1){$\epsilon$} &
\node (i) {$\dots$}; &
\inetcell(n){$\epsilon$} \\ };
\inetwirefree(1.pal)
\inetwirefree(n.pal)
\end{tikzpicture}
$$
\end{subfigure}
\hfill
\begin{subfigure}[t]{0.4\textwidth}
\centering
\caption{Duplication}
\label{egdupfig}
$$
\begin{tikzpicture}[baseline=(a)]
\matrix[row sep=1em]{
\inetcell(d){$\delta$}[D] \\
\inetcell(a){$\phantom x\alpha\phantom x$}[U] \\ };
\node (i) [below=of a.middle pax] {$\dots$};
\inetwirefree(a.left pax)
\inetwirefree(a.right pax)
\inetwirefree(d.left pax)
\inetwirefree(d.right pax)
\inetwire(d.pal)(a.pal)
\end{tikzpicture}
\rightarrow
\begin{tikzpicture}[baseline=(i.base)]
\matrix[row sep=2em]{
\inetcell(l){$\phantom x\alpha\phantom x$}[U] & &
\inetcell(r){$\phantom x\alpha\phantom x$}[U] \\
\inetcell(1){$\delta$} &
\node (i) {$\dots$}; &
\inetcell(n){$\delta$} \\ };
\node (li) [below=of l.middle pax] {$\dots$};
\node (ri) [below=of r.middle pax] {$\dots$};
\inetwire(l.left pax)(1.right pax)
\inetwire(l.right pax)(n.right pax)
\inetwire(r.left pax)(1.left pax)
\inetwire(r.right pax)(n.left pax)
\inetwirefree(1.pal)
\inetwirefree(n.pal)
\inetwirefree(l.pal)
\inetwirefree(r.pal)
\end{tikzpicture}
$$
\end{subfigure}
\hfill
\begin{subfigure}[t]{0.2\textwidth}
\centering
\caption{Non-termination}
\label{egloopfig}
$$
\begin{tikzpicture}[baseline=(g.above pal)]
\matrix[row sep=0.7em]{
\inetcell(g){$\gamma$}[D] &
\inetcell[opacity=0](gr){$\gamma$}[D] \\
\inetcell(d){$\delta$}[U] &
\inetcell[opacity=0](dr){$\delta$}[U] \\
};
\inetcell[above=of g.above right pax](et){$\epsilon$}[D]
\inetcell[below=of d.above left pax](eb){$\epsilon$}[U]
\inetwire(g.pal)(d.pal)
\inetwire(et.pal)(g.right pax)
\inetwire(g.left pax)(gr.middle pax)
\inetwire(d.right pax)(dr.middle pax)
\inetwire(eb.pal)(d.left pax)
\inetwirecoords(gr.middle pax)(dr.middle pax)
\end{tikzpicture}
\righttoleftarrow^*
$$
\end{subfigure}
\end{figure}

The interaction calculus defines reduction on configurations in more details than seen from graph rewriting defined on interaction nets.
Namely, if $\alpha[v_1, \dots, v_m] \bowtie \beta[w_1, \dots, w_n]$, the following reduction:
$$
\langle\alpha(t_1,\dots, t_m) = \beta(u_1,\dots, u_n),\dots\rangle
\rightarrow
\langle t_1 = v_1,\dots,\ t_m = v_m,\ u_1 = w_1,\dots,\ u_n = w_n,\dots\rangle
$$
is called \textit{interaction}.
For equations of the form $x = u$ \textit{indirection} can be applied resulting in \textit{substitution} $t[x := u]$ defined as the result of replacing the other occurrence of the name $x$ in term $t$ with term $u$:
$$
\langle x = u,\ t = w,\dots\rangle
\rightarrow
\langle t[x := u] = w,\dots\rangle.
$$
An equation $t = x$ is called a \textit{deadlock} if the name $x$ has occurrence in the term $t$.
Together, interaction and indirection define the reduction relation on configurations.

Coming back to the example of a non-terminating net shown in Figure~\ref{egloopfig}, the infinite reduction sequence starting from the corresponding configuration in the interaction calculus is as follows:
\begin{align*}
&\langle\delta(\epsilon, x) = \gamma(x, \epsilon)\rangle \rightarrow \\
&\langle\epsilon = \gamma(x_1, x_2),\ x = \gamma(y_1, y_2),\ x = \delta(x_1, y_1),\ \epsilon = \delta(x_2, y_2)\rangle \rightarrow^* \\
&\langle x_1 = \epsilon,\ x_2 = \epsilon,\ x = \gamma(y_1, y_2),\ x = \delta(x_1, y_1),\ x_2 = \epsilon,\ y_2 = \epsilon\rangle \rightarrow^* \\
&\langle\delta(\epsilon, x) = \gamma(x, \epsilon)\rangle \rightarrow \dots
\end{align*}

Now, let us describe the properties of interaction nets more formally:
\begin{itemize}
\item \textit{locality} means that only active pairs can be rewritten;
\item \textit{linearity} means that each interaction rule can be applied in constant time;
\item \textit{strong confluence} also known as \textit{one-step diamond property} means that if $c \rightarrow c_1$ and $c \rightarrow c_2$, then $c_1 \rightarrow c'$ and $c_2 \rightarrow c'$ for some $c'$.
\end{itemize}

Perhaps the simplest universal interaction system is that of interaction combinators~\cite{comb}.
It is defined by the signature $\Sigma = \{\epsilon, \delta, \gamma\}$ with annihilation rules $\gamma[x, y] \bowtie \gamma[y, x]$ and $\delta[x, y] \bowtie \delta[x, y]$ in addition to the erasing and duplication rules shown above.
The interaction system of interaction combinators is Turing-complete and can simulate any other interaction system or another model of computation.

However, please note that signatures and sets of interaction rules are not required to be small or even finite.
In fact, in many applications of interaction nets infinite sets of agent types and interaction rules are actually preferred as they often allow to represent desired structures in a much more natural and efficient way than using interaction combinators.

\section{Conclusion}

In this note we briefly presented interaction nets, which we believe may be a worthwhile replacement for cellular automata to consider in digital physics.
It would be interesting to see the consequences of using clockless computation with graph-like structures as the basis for discrete, deterministic characterisation of the physical world, in particular with respect to the phenomenon of quantum entanglement.

\nocite{*}
\bibliographystyle{eptcs}
\bibliography{cite}
\end{document}